1# Deep Learning Enables Robust and Precise Light Focusing on Treatment Needs

Changchun Yang, Hengrong Lan, and Fei Gao, *Member, IEEE*

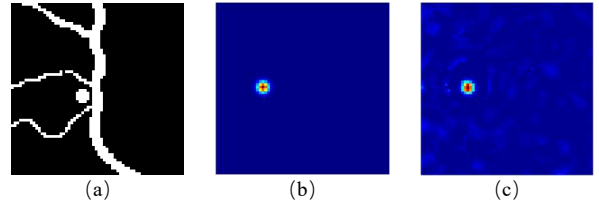

Figure 1: (a) An example of a medical image after tomography imaging and segmentation, containing the blood vessel and the tumor. (b) Ground truth $y_f$ is the area of treatment concern, we want all the light to be focused on the tumor, rather than evenly escaping to the entire slice, thereby achieving an energy enhancement. (c) The focus result $\widetilde{y}_f$ after transmission process obtained by phase $\widetilde{x}_f$, which is predicted by LoftGAN that hope to form the optical focusing (b). The scale range of (b) and (c) is 0~1.

*Abstract—* If light passes through the body tissues, focusing only on areas where treatment needs, such as tumors, will revolutionize many biomedical imaging and therapy technologies. So how to focus light through deep inhomogeneous tissues overcoming scattering is Holy Grail in biomedical areas. In this paper, we use deep learning to learn and accelerate the process of phase pre-compensation using wavefront shaping. We present an approach (LoftGAN, light only focuses on treatment needs) for learning the relationship between phase domain $X$ and speckle domain $Y$. Our goal is not just to learn an inverse mapping $F: Y \to X$ such that we can know the corresponding $X$ needed for imaging $Y$ like most work, but also to make focusing that is susceptible to disturbances more robust and precise by ensuring that the phase obtained can be forward mapped back to speckle. So we introduce different constraints to enforce $F(Y) \approx X$ and $H(F(Y)) \approx Y$ with the transmission mapping $H: X \to Y$. Both simulation and physical experiments are performed to investigate the effects of light focusing to demonstrate the effectiveness of our method and comparative experiments prove the crucial improvement of robustness and precision. Codes are available at https://github.com/Changchun-Yang/LoftGAN.

## I. Introduction

If the disordered medium is thicker than some mean free path (about 0.1 mm inside living tissue), multiple scattering occurs after the light enters it due to the refractive index mismatch [1]. A large amount of information carried by incident light will be lost due to scattering. This is highly detrimental to high-resolution optical transmission and imaging through or in biological tissue. When the coherent light scatters, it will randomly interfere to form speckles. Although the speckle appears to be completely and randomly formed, the process of the scattering actually has been proven to be deterministic at a particular time window (speckle correlation time [2, 3, 29]). In this way, the wavefront shaping [4] is proposed for optical focusing and imaging in strongly scattering medium, which uses a spatial light modulator (SLM) to control optical wavefront to reduce the limitation of image resolution by light scattering in disordered media.

Many novel medical imaging methods are inseparable from light, such as photoacoustic tomography [5], diffuse optical tomography [6]. When photons are randomly scattered, the tomography of the whole slice will be more evenly presented, but often the treatment will pay more attention to the abnormal areas, such as the tumor (Figure 1(a)). So when getting a complete image of the imaging results, we can firstly segment the tumor, and hope that the photons can focus as much as possible on this area (Figure 1(b)) to achieve effective treatment noninvasively.

In wavefront shaping, the propagation of light emitted by the laser is controlled by the free space light propagation theory, which will be affected by the phase $x$ on SLM's screen when passes through SLM, and then scatters through the medium to form corresponding speckle, the intensity $y$ of which is recorded by a camera, $H$ denotes the forward scattering function (Figure 2(a)). The core of the work is to find the cycle relationship between phase domain $X$ and the speckle domain $Y$, and focus the light on the abnormal area. In recent years, there have been many remarkable achievements in segmenting abnormal areas such as tumors [7-9], with these methods we

Changchun Yang, Hengrong Lan and Fei Gao are with the Hybrid Imaging System Laboratory, Shanghai Engineering Research Center of Intelligent Vision and Imaging, School of Information Science and Technology, ShanghaiTech University, Shanghai 201210, China (*corresponding author: gaofei@shanghaitech.edu.cn).

Changchun Yang and Hengrong Lan are also with the Chinese Academy of Sciences, Shanghai Institute of Microsystem and Information Technology, Shanghai 200050, China, and University of Chinese Academy of Sciences, Beijing 100049, China.

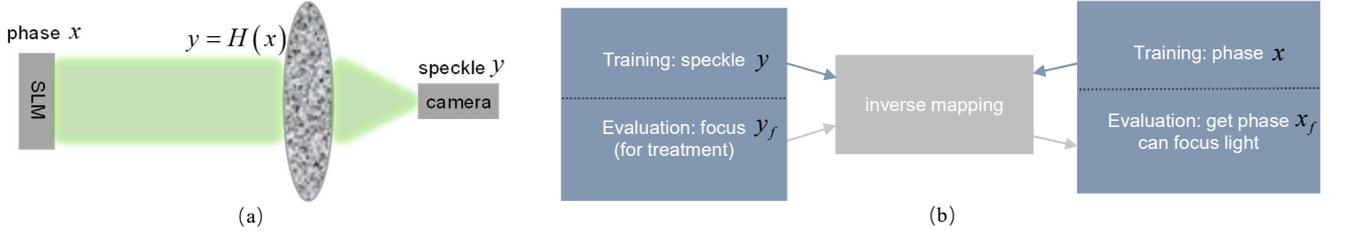

Figure 2: (a) Incident light passing through the SLM with different phases will cause different speckles after multiple scattering in an inhomogeneous medium. (b) When using neural networks to solve the inverse mapping $F$, paired speckle domain $Y$ and phase domain $X$ are used in training, where $F$ is the input. And region of treatment attention is used as input for evaluation to obtain the corresponding phase.

can achieve accurate medical image segmentation and get the area we want the light to focus on, hence the key of this paper is to find the SLM phase $x_f$ that can make the light precisely reach the focus effect.

Due to the complexity of the transmission matrix and the existing wavefront shaping method, it usually takes large amount of time and computational resources to find the best input mode that can achieve light focusing after passing through a diffuser. Machine learning, especially deep learning, has shown superior performance over traditional models in solving the inverse problem, such as denoising [10, 11], super resolution [12], and image reconstruction [13, 14]. Using neural networks to explore light focusing are also beginning to appear in very recent years [15-17]. However, the current methods are mostly very simple, and only briefly consider the estimation of the inverse function without the essence of the problem of optical focusing. These methods are simple fitting reverse mapping without any constraints. Based on this, the results obtained cannot meet the needs of treatment.

In this work, we investigate how to learn the inverse mapping $F$, and find the corresponding $X$ from $Y$ (Figure 2(b)). Besides, we hope to further obtain $Y$ through the predicted $X$, so as to learn the forward mapping $H$, to verify the identity $H \circ F = I$. In a sense, just like we translate a sentence from English to German and then translate it back from German to English, we should go back to the original sentence. We design a network that combines reverse-forward, named LoftGAN, for more robust and precise light focusing. More specifically, we introduce deviation loss and distribution loss to enhance $F(Y) = \widetilde{X} \approx X$, content loss and style loss to enhance $H(F(Y)) = \widetilde{Y} \approx Y$, respectively. We also explore the possibility of achieving complex focus, such as multiple points, and compares the robustness and precision of the focusing results. Simulation and physical experiments are demonstrated to demonstrate the superiority of our focus.

## II. RELATED WORK

**Wavefront Shaping** is based on the transmission matrix [18, 19] to modulate the phase before the incident light enters the scattering medium. In this way, the phase distortion caused by the scattering is pre-compensated. Therefore, many algorithms for achieving optimal compensation have been proposed, which greatly accelerate the development of wavefront shaping, such as continuous sequential algorithm [20], genetic algorithm [21], or phase-conjugation of the transmission matrix [5]. However, these algorithms have obvious limitations like time and resource consuming. Many linear operations or nonlinear operations of the optimization algorithm suggest that the neural network may be a good strategy for this field. Deep learning assisted wavefront shaping can learn the complex relationship, currently in preliminary exploration. But this is undoubtedly a promising direction to solve optical focusing in scattering medium. The relative research of using deep learning to assist wavefront shaping has also been introduced in the first section. However, these methods are often limited to simple several layers of neural networks. The precision and robustness of the focusing results without other constraints on a single learning reverse mapping cannot meet the treatment needs.

**AutoEncoder** is a powerful model that can be used for dimensionality reduction or feature extraction [22]. Due to its flexibility and extensibility, it is now also used in the generative model, e.g. Variational AutoEncoder [23]. Its encoder compresses the input, encodes it as a code, and then decoder decodes the code into input. Inspired by this, we encode a speckle $y$ as the phase $\widetilde{x}$ and then decode $\widetilde{x}$ to $\widetilde{y}$. The difference is that we use $x$ to supervise the encoded code $\widetilde{x}$, to guarantee the deviation and distribution between them. Our encoder and decoder are completely symmetrical, so as to learn the inverse mapping function and the forward mapping function, respectively.

**Generative Adversarial Networks (GANs)** [24] continues to play through a generator and discriminator until it reaches the Nash equilibrium, which produces good results in many problems. GANs use the generator to map latent $z$ to data space, which will be judged by the discriminator whether it is a generated sample or a real sample. We use GAN's generator to translate the phase $\widetilde{x}$ to $\widetilde{y}$ instead of the decoder of AutoEncoder by sharing parameters, and then use a discriminator. Besides, we also use this generator to encode ground truth $x$ to $\hat{y}$, so the style loss [25] among $y, \widetilde{y}, \hat{y}$ is used to produce better focusing results. In order to ensure the details of our method, we use the PatchGAN discriminator [26] to output a $N \times N$ patch, which has fewer parameters and can handle images of any size using full convolution. We also introduce the content loss by using the hidden layer representation of the discriminator between $y$ and $\widetilde{y}$, to constrain the encoder and the generator.

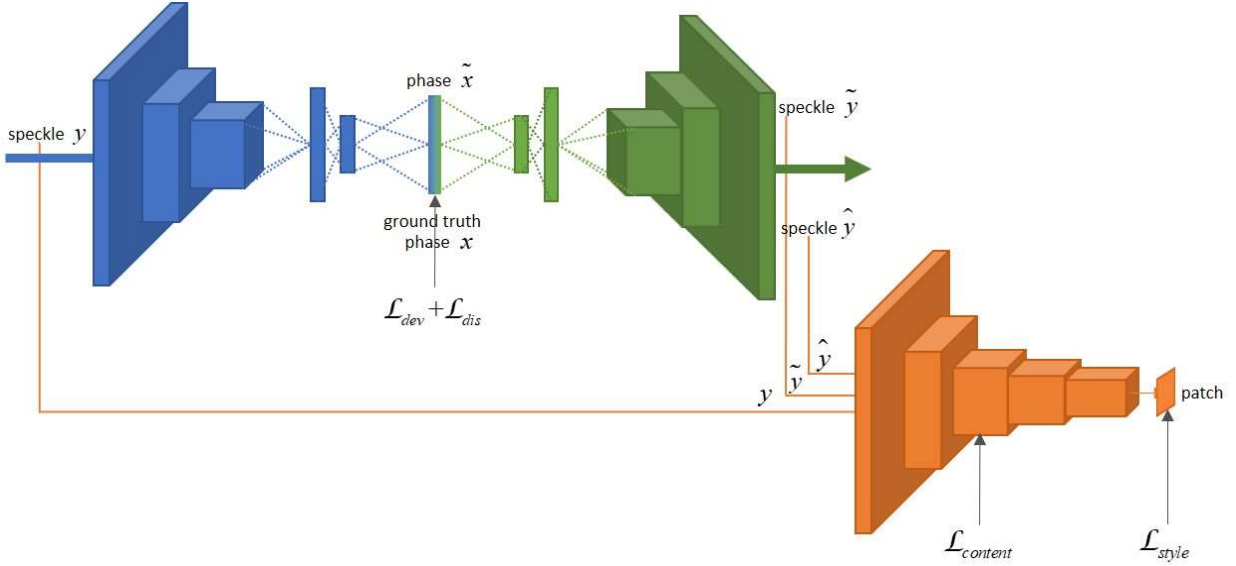

Figure 3: Our model contains three parts, the encoder, generator, and discriminator, expressed in blue, green, and orange, respectively. The encoder part $Enc$ convert the input (speckle $y$) to the phase domain by learning the inverse mapping function $F$ to get the phase $\tilde{x}$, and the deviation loss $\mathcal{L}_{dev}$ and distribution loss $\mathcal{L}_{dis}$ are used as constraints for $\tilde{x}$ from values and distributions to be closer to the ground truth $x$. The phase $\tilde{x}$ and $x$ are each used as input (share parameters) to the generator part $Gen$ and then get their respective outputs: $\tilde{y}$ and $\hat{y}$, together with $y$ will be sent to the discriminator part $Dis$. The content loss $\mathcal{L}_{content}$ and the style loss $\mathcal{L}_{style}$ are used in the hidden layer and the last patch layer.

## III. METHODOLOGY

Our goal is to learn the inverse mapping function $F$ between speckle domain $Y$ and phase domain $X$ giving training samples, the speckle $\{y_i\}_{i=1}^{N}$ where $y_i \in Y$ and the phase $\{x_i\}_{i=1}^{N}$ where $x_i \in X$. Then giving the focus area $y_f$ of treatment attention, we can get the corresponding $\widetilde{x_f}$ so that the light can pass through the scattering medium and focus as much as possible on the area instead of scattering, i.e. $H(\widetilde{x_f}) \approx y_f$. We introduce different constraints to supervise the reverse and forward learning process. As illustrated in Figure 3, overall objective contains four types of terms relying on different metrics, whose details based on our practical considerations about the optical focusing will be introduced next. Combining with Algorithm 1, an overview of the training process is given.

### A. Phase Loss

Predicting the phase involves two losses $\mathcal{L}_{dev}$ and $\mathcal{L}_{dis}$, the $Enc$ encoder a speckle sample $y$ to a phase representation $\tilde{x}$, and the $Gen$ decode the phase representation back to speckle domain:

$$\tilde{x} \sim Enc(y) = q(\tilde{x}|y), \tilde{y} \sim Gen(\tilde{x}) = p(y|\tilde{x}). \quad (1)$$

We regularize the result of predicting phase by imposing two constrains, a F-norm over the ground truth phase $x$ and a prior regularization over the ground truth distribution $p(x)$:

$$\mathcal{L}_{dev} = \|\tilde{x} - x\|_F^2 \quad (2)$$

$$\mathcal{L}_{dis} = D_{KL}\left(q(\tilde{x}|y) \| p(x)\right) \quad (3)$$

where $D_{KL}$ is the Kullback-Leibler divergence.

### B. Adversarial Loss

The GANs' goal is to find a binary classifier that can make the best discrimination between real sample data and generated data and simultaneously expect $Gen$ to fit the real sample distribution. And the traditional adversarial loss [24] is used to minimize the binary cross entropy:

$$\mathcal{L}_{GAN} = \mathbb{E}_x\left[\log(D(x))\right] + \mathbb{E}_z\left[\log(1 - D(G(z)))\right] \quad (4)$$

Note that the regularization of the phase domain $\mathcal{L}_{dis}$ should make the speckle data from either $q(\tilde{x}|y)$ or $p(x)$ be similar. However, when we input an any speckle sample $y$ for the network, the generated data $Gen(Enc(y))$ is much more likely to be similar to $y$ than $Gen(x)$. Moreover, receiving positive and negative samples at the same time helps to learn better, hence we introduce our style loss:

$$\begin{aligned}\mathcal{L}_{style} = & \mathbb{E}_y\left[\log(Dis(y))\right] \\ & + \mathbb{E}_x\left[\log(1 - Dis(Gen(x)))\right] \\ & + \mathbb{E}_y\left[\log(1 - Dis(Gen(Enc(y))))\right]\end{aligned} \quad (5)$$

### C. Content Loss

One advantage of GAN is that its discriminator can implicitly learn the rich similarity metric of the images in order to distinguish between real and fake images. Therefore, this observation can be utilized, so that the image characteristics learned by the discriminator can be converted into more abstract reconstruction error [27] after $Enc$ and $Gen$. This can be achieved by denoting the hidden representation of the $lth$ layer of the discriminator $Dis$ by $Dis_l(y)$. Since we hope to get $\tilde{x}$ similar to $x$ through the $F$ learned by $Enc$, we combine with the $H$ learned by $Gen$, i.e. $H \circ F$ to get $\tilde{y}$, which also should be similar to a $y$. Hence a gaussian observation model for $Dis_l(y)$ with mean $Dis_l(\tilde{y})$ and identity covariance is introduced:

$$p(Dis_l(y)|\tilde{x}) = \mathcal{N}(Dis_l(y)|Dis_l(\tilde{y}), I) \quad (6)$$

We can get the content loss between $y$ and $\tilde{y}$:

$$\mathcal{L}_{content} = -\mathbb{E}_{q(\tilde{x}|y)}\left[\log p(Dis_l(y)|\tilde{x})\right] \quad (7)$$

*D. Full Objective*

Combined with phase loss, adversarial loss and content loss, our full objective is:

$$\mathcal{L} = \mathcal{L}_{dev} + \mathcal{L}_{dis} + \mathcal{L}_{content} + \mathcal{L}_{style} \quad (8)$$

which is the goal we are going to optimize.

*E. Limiting Loss Signals*

We can train the three parts, encoder, generator, and discriminator simultaneously by using the combined loss function in Eq. 8 and not updating all the network parameters with respect to $\mathcal{L}$ during training. More specifically, $Dis$ should not try to minimize $\mathcal{L}_{content}$ because this would collapse $Enc$ to 0, and the loss signal from $\mathcal{L}_{style}$ should not be backpropagated to $Enc$. We do not set weights for each of the four loss terms of the full objective (Eq. 8), but when updating the parameters, we still apply different hyperparameters to different parts. As $Enc$ receives a loss signal from three terms $\mathcal{L}_{dev}$, $\mathcal{L}_{dis}$, and $\mathcal{L}_{content}$, three weights are used here to measure the ability of each. Two terms $\mathcal{L}_{content}$ and $\mathcal{L}_{style}$ are applied to $Gen$, which are also assigned a weight.

## IV. IMPLEMENTATION

The structure of the proposed LoftGAN is shown in Figure 3, which has three parts encoder, generator, and discriminator. U-Net [28] demonstrates its powerful effects in many fields by combining low-level semantic information with high-level fine features. It extracts image features in the contracted path and maps the features back to the original image resolution on the extended path. So we use a completely symmetrical structure in the encoder and generator to learn the inverse mapping function $F$ and the forward mapping function $H$. For the encoder, it contains three convolution layers and two full connection layers. 16, 32, 48 filters are used in the three convolution layers, and the filter size of each layer is 7×7, 5×5, 3×3, and their strides are set as 3×3, 2×2, 1×1, respectively. The first full connection

---

**Algorithm 1** Training the LoftGAN model

$\theta_{Enc}$, $\theta_{Dec}$, $\theta_{Dis}$ ← initialize network parameters
**repeat**
$\quad Y$ ← random mini-batch from speckle domain
$\quad \tilde{X} \leftarrow Enc(Y)$
$\quad \mathcal{L}_{dev} \leftarrow \|\tilde{X} - X\|_F^2$
$\quad \mathcal{L}_{dis} \leftarrow D_{KL}(q(\tilde{x}|y) \| p(x))$
$\quad \tilde{Y} \leftarrow Gen(\tilde{X})$
$\quad \mathcal{L}_{content} \leftarrow -\mathbb{E}_{q(\tilde{x}|y)}\left[\log p(Dis_l(Y)|\tilde{X})\right]$
$\quad X$ ← mini-batch from phase domain corresponding to $Y$
$\quad \hat{Y} \leftarrow Gen(X)$
$\quad \mathcal{L}_{style} \leftarrow \mathbb{E}_Y[\log(Dis(Y))] + \mathbb{E}_{\tilde{Y}}[\log(1 - Dis(\tilde{Y}))]$
$\quad\quad\quad\quad\quad + \mathbb{E}_{\hat{Y}}[\log(1 - Dis(\hat{Y}))]$
$\quad$ // Update parameters using gradient descent
$\quad \theta_{Enc} \stackrel{+}{\leftarrow} -\nabla_{\theta_{Enc}}(\lambda_{dev}\mathcal{L}_{dev} + \lambda_{dis}\mathcal{L}_{dis} + \lambda_{con}\mathcal{L}_{content})$
$\quad \theta_{Gen} \stackrel{+}{\leftarrow} -\nabla_{\theta_{Gen}}(\lambda'_{con}\mathcal{L}_{content} - \mathcal{L}_{style})$
$\quad \theta_{Dis} \stackrel{+}{\leftarrow} -\nabla_{\theta_{Dis}}\mathcal{L}_{style}$
**until** death

---

layer has 512 neurons followed by a dropout layer, whose dropout rate is 0.5. And the number of neurons in the second full connection layer is the same as the size of the phase. Symmetrically, the generator has two full connection layers and three convolution layers.

As for the discriminator, it firstly uses three convolution layers identical to the encoder. The output at this time is used to calculate the content loss for similarity measurement. Finally, three convolutional layers are used to reduce the number of channels to 1, and output a N×N patch. Each element of the matrix is output 0 or 1 by the discriminator. We use keras to build the entire LoftGAN model, and one GPU (NVIDIA RTX Titan) is used for computing. And the four hyperparameters in Algorithm 1 are 22, 0.03, 0.03 and 1e-6 in order.

## V. EXPERIMENT

We perform simulation experiments and physical experiments to collect phase domain data and corresponding speckle domain data, where ground truth speckle-phase pairs are available for training the LoftGAN. The abnormal area, such as tumor is used for evaluation, in order to obtain the phase that can achieve optical focusing to the area of the treatment target. In this work, we use speckles of size 64×64, phases of size 32×32 (1024). The details of the simulation transmission process and the physical experiment setup are given below. We then conduct ablation studies to study the importance of each module in our approach, and demonstrate their role in achieving optical focusing by comparing their performance and in-depth analysis.

*A. Simulation Experiment*

We firstly use simulations close to the physical model to verify the possibility of optical focusing. A M×N transmission matrix $t$ is generated to represent the light scattering path in the disordered medium. The transmission matrix $t$ follows the circularly-symmetric complex normal distribution [21]. So, first of all, a random matrix following this distribution is generated. Then a set of Hadamard basis are generated as input basis. After matrices multiplication, the electric fields obtained propagate to far field. The propagation process is governed by the free space light propagation theory. Then the electric fields reaching the receiving plane together with input Hadamard basis is used to calculate the whole transmission matrix.

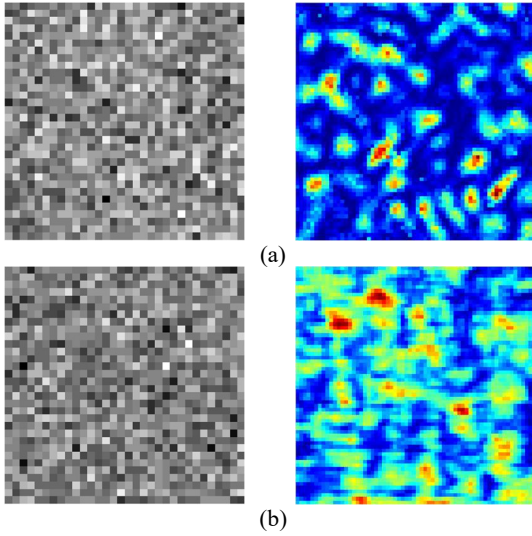

Figure 4: The results of one pair of simulation example (a) and experimental example (b) are given respectively. The left side shows random phases, and the right side is the corresponding speckle pattern formed by scattering in the inhomogeneous tissue after the light carries the left phase information.

After obtaining the transmission matrix $t$, a set of random phase patterns $X$ are generated to multiply with $t$ to obtain different speckle patterns $Y$ for LoftGAN training (Figure 4(a)). Specifically, after getting $t$, we can model the scattering process as a linear relation between the complex field at the input modes and the complex field at the output modes [21, 29], that is, the incident light and the transmitted light are coupled through $t$:

$$E_m = \sum_{n}^{N} t_{mn} A_n e^{i\phi_n} \quad (9)$$

where $\phi_n$ is the phase at input mode n, corresponding to the $n$-th position of the phase $x \in X$ we generated. $E_m$ is the complex field at output mode m. The amplitude contribution from input mode n is represented by $A_n$. Assuming that the input plane is illuminated homogeneously, all incoming modes have the same intensity, so $A_n$ can be defined as: $A_n = 1/\sqrt{N}$. $t_{mn}$ relates the field at input mode n to the field at output mode m, which is the element of the transmission matrix $t$.

Achieving optical focus at a certain area is essentially to maximize the intensity of some specified output modes, and the intensity transmitted into mode m is given by

$$I_m = \frac{1}{N}\left|\sum_{n}^{N} t_{mn} e^{i\phi_n}\right|^2 \quad (10)$$

which will also be recorded by the CCD camera. Based on these, we get 12888 pairs speckle-phase, which will be used as samples for training LoftGAN. We enter $y_f$ into the trained model during verification, get the output phase $\widetilde{x_f}$, which will be treated as the input phase in Eq. 9, then get the final focus test results in Figure 1(c).

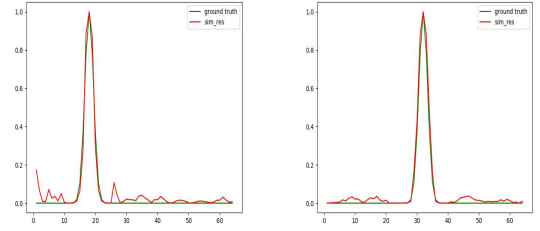

Figure 5: Compare the focus result Figure 1(c) with ground truth Figure 1(b). The left side shows light intensity profile along horizontal direction through focal point. And the right side shows light intensity profile along vertical direction through focal point.

We also compared the cross-sectional intensity distribution at the focal point, as shown in Figure 5. It can be analyzed that our simulation results are almost perfectly focused in the entire target area (round tumor). Although there is escape light in other positions (ideally a very small number close to 0), when the intensity is low, it will not harm normal human tissues and meet the needs of treatment.

### B. Physical Experiment

The physical experimental setup is illustrated in Figure 6. The resolution of the SLM screen is 1024×1920, and we divide it to 32×32 macro-pixels to display the phase $x \in X$, so one macro-pixel contains 32×60 pixels. And then a camera is used to record the speckle $y \in Y$ with the size of 64×64 pixels. 32 grey levels are used in the SLM screen to represent the phase values from 0 to $2\pi$.

A continuous wavelength laser beam ($\lambda = 670nm$) firstly is expanded by 10 times through two lenses ($f_{L1} = 15mm$, $f_{L2} = 150mm$), adjusting the incident light to a suitable polarization state using a polarizer before transmitting to the liquid crystal spatial light modulator. Two additional lenses ($f_{L3} = 200mm$, $f_{L4} = 50mm$) are placed behind the SLM. We then place two objectives (OB1, OB2: 10X), one focuses the light before entering the scattering medium to achieve better scattering in the scattering medium, and the other one collects the scattered light. Finally, using a CCD camera to record the intensity information of the emitted light, we can get the speckle samples we need, which will be used as inputs for training the LoftGAN.

We collected the same number of samples as the simulation experiment. Before we use the experimental data to train the LoftGAN, we need to preprocess it and normalize the obtained speckles and phases to 0~1. After the trained model is obtained, the tumors (target focus results), such as the four ground truths in Figure 7, are sent to the LoftGAN, predicting the output phases, which are then loaded onto the SLM screen to adjust the incoming optical wavefront. The final focus results using the predicted phases are recorded.

Table 1: Experimental device detail

| Laser | 670nm, MRL-F671-1W CW |
|---|---|
| SLM | HDSLM80R |
| Camera | TUCAM |
| Diffuser | GCL-200101 |
| Objective Lens | GCO-2102, GCO-2105 |

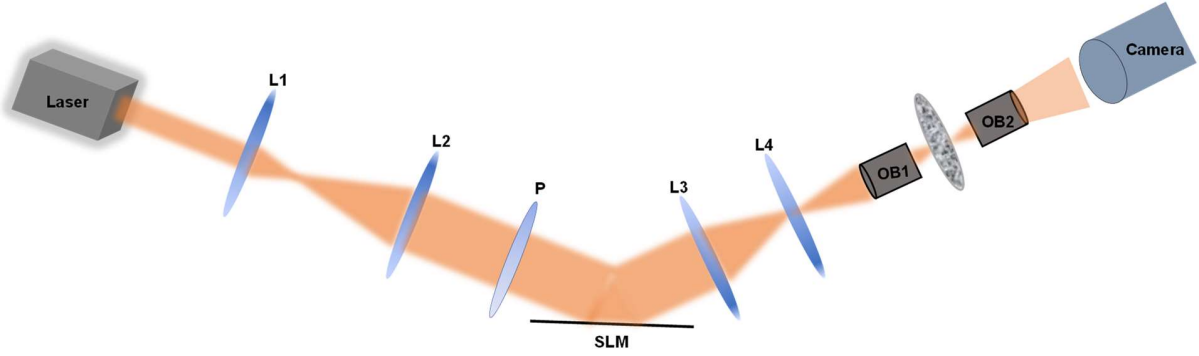

Figure 6: Schematic of our physical experimental setup. We first use two lenses (L1 and L2) to expand the light from the laser, and a polarizer (P) is used to adjust the polarization of incident light into the spatial light modulator (SLM). The light is reflected after being modulated by the input mode on the SLM screen, then passes through the two lenses (L3 and L4), then we use a microscope objective (OB1) to demagnify and image the SLM. OB1 focuses the light beam through the scatter (D), then a second microscope objective (OB2) is used to collect the scattered light, and light intensity is recorded by a CCD camera.

More detailed configuration information about the experiment setup is shown in Table 1. And comparison experimental results are given in Figure 8.

As shown, we further design the following three approaches for comparsion:
1) Only use the encoder $Enc$ with $\mathcal{L}_{dev}$ and $\mathcal{L}_{dis}$
2) Remove $\mathcal{L}_{dis}$, that is LoftGAN without the $\mathcal{L}_{dis}$ denoted as $LoftGAN - \mathcal{L}_{dis}$
3) Remove $\mathcal{L}_{content}$, that is LoftGAN without the $\mathcal{L}_{content}$ denoted as $LoftGAN - \mathcal{L}_{content}$

As reflected from the results, the use of generator $Gen$ and discriminator $Dis$ is necessary, this can be found by observing the focus results only for the encoder $Enc$. Since we only focus on the inverse mapping function, without the help of estimating the forward mapping function, the whole learning process is particularly unstable. When achieving the optical focusing of multiple points, such as the two points in the example, the point on the right is almost a failure because the intensity is very low. Then we remove the loss term $\mathcal{L}_{dis}$, and the difference of the focused visual effect between $LoftGAN - \mathcal{L}_{dis}$ and the LoftGAN is small, but by measuring the numerical intensity of the light intensity, we find that the variance is greater than the LoftGAN, so it is meaningful to constrain the distribution of the desired phase. $\mathcal{L}_{content}$ is used to enforce $H(F(Y)) \approx Y$ by using the hidden output of the discriminator, will give a deep supervisory on the training of the encoder and generator. We demonstrate our method can achieve specific precision optical focusing through scatter tissues.

## VI. DISCUSSION AND CONCLUSION

Optical focusing plays a significant role in many biomedical applications, which is clinically meaningful and instructive. By focusing the light, we can increase the energy enhancement at specific locations, so that there are many strategic applications such as non-invasive treatment or deep tissue imaging. This will drive advances in many light-based imaging methods, but because light undergoes multiple scattering as it passes through living tissues, making it very challenging to achieve optical focusing deep in the tissues. In recent years, with the advancement of wavefront shaping, there have been some work reporting that making light focus tightly and efficiently through or in thick and inhomogeneous media. These methods can be reduced to two categories, one is based on optimization algorithms, which are very time consuming and computationally intensive, and the placement experiment is very complicated. The other type is based on deep learning, which is only in its infant stage. The reported experimental results in very limited literatures are all based on single point focusing, and their network used is too rudimentary, such as single-layer fully connected network. The intrinsic analysis of combined neural network-assisted optical focusing has not yet appeared.

For the fist time to the best of our knowledge, starting from the inverse mapping function $F$ and the forward mapping function $H$ of the fundamental transmission process, we implement a symmetric encoder and generator of GAN, and design a discriminator to identify the three results: symmetric coupling result of encoder and generator, the output of the generator acting alone phase and the speckle of the input. Also we have deep supervision of intermediate results, verified by ablation studying. Our method can clearly, efficiently and accurately give the phase that can focus on the abnormal area.

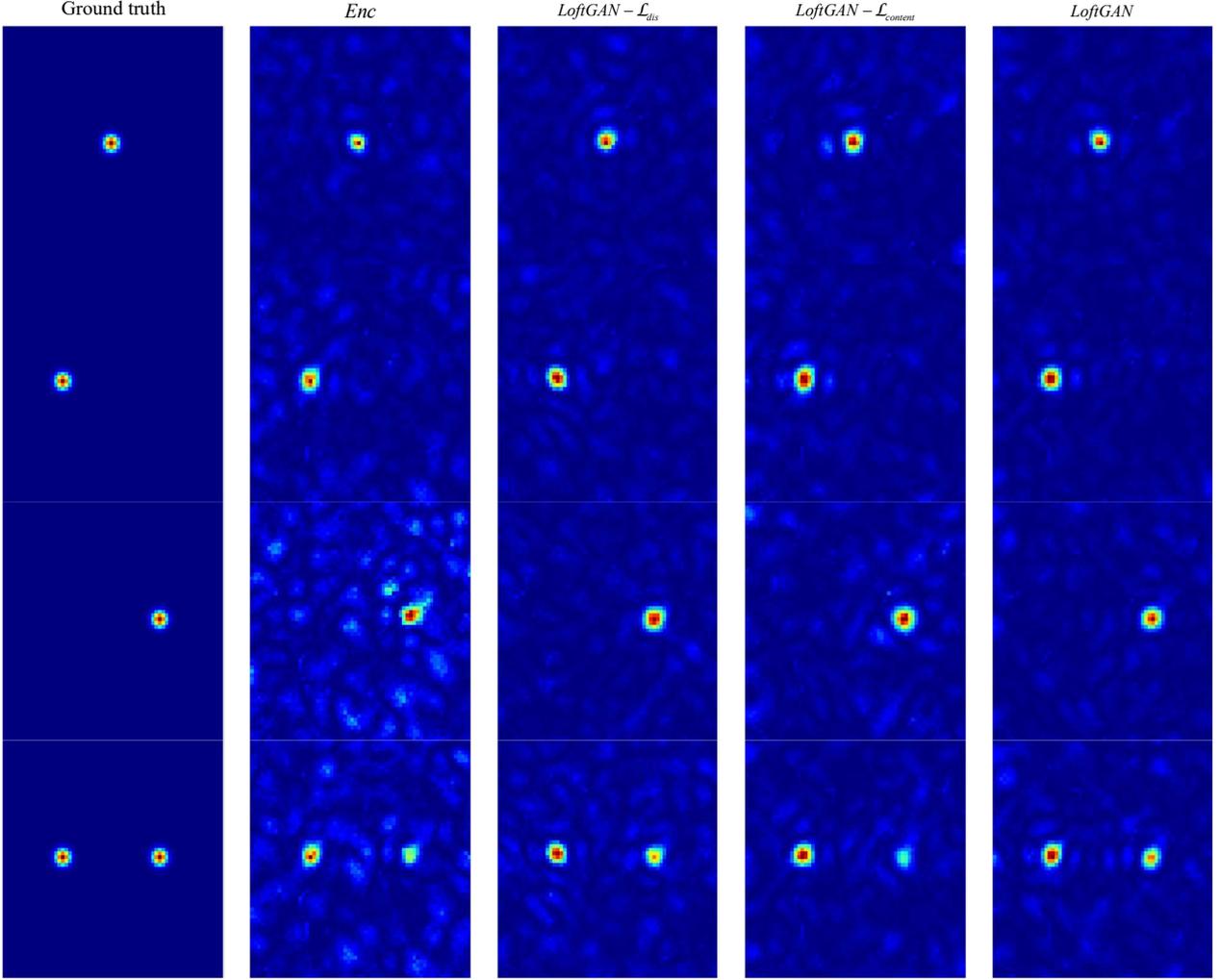

Figure 7: Do ablation learning. Ground truth is shown on the leftmost column, then we only use the encoder $Enc$ to get the phases, then use the phases to get the focusing results, which are display on the second column. And so on, the results of LoftGAN without the $\mathcal{L}_{dis}$, without the $\mathcal{L}_{content}$, and complete LoftGAN are shown on the next three columns. The scale range of all images is 0~1.

Our approach has a common limitation same with other types of neural networks: limited by the number of training samples. So, using more samples may lead to better focus, but it takes more time and memory to collect the samples. This trade-off between performance and time consumption deserves careful consideration in such application. At the same time, we have not studied whether it can accurately focus after increasing the disturbance in the actual environment. In the future work, we will more fully verify the optical focusing in various environments, especially in the light-related imaging, to explore the possibility of more flexible focusing. In the long term, we expect our proposed deep-learning based optical focusing method will significantly advance the biomedical area, achieving deep-penetration high-resolution imaging, as well as non-invasive and precise cancer treatment.

ACKNOWLEDGMENT

This research was funded by Start-up grant of ShanghaiTech University (F-0203-17-004), Natural Science Foundation of Shanghai (18ZR1425000), and National Natural Science Foundation of China (61805139). The data generation of the simulation experiment and the setup of the physical experiment both have received guidance from Yunqi Luo. Thanks for her support.

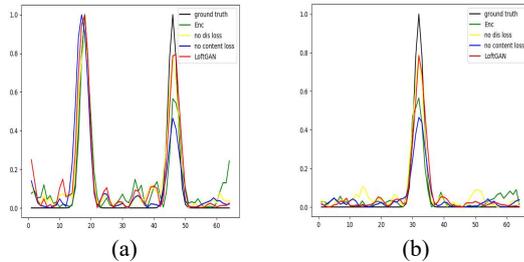

Figure 8: Compare the intensity of the focus effects of the four methods. (a) Focusing the light into two points, compare the intermediate horizontal light intensity. It is obvious that the focus effect of the point on the right side is best when using LoftGAN. (b) Vertical horizontal light intensity at the right point.

REFERENCES

[1] J. Bertolotti, E. G. van Putten, C. Blum, A. Lagendijk, W. L. Vos, and A. P. Mosk, "Non-invasive imaging through opaque scattering layers," *Nature,* vol. 491, no. 7423, pp. 232-234, 2012/11/01 2012.

[2] J. E. Fernández and V. Scot, "Deterministic and Monte Carlo codes for multiple scattering photon transport," *Applied Radiation and Isotopes,* vol. 70, no. 3, pp. 550-555, 2012/03/01/ 2012.

[3] Y. Luo, S. Yan, H. Li, P. Lai, and Y. Zheng, "Towards smart optical focusing: Deep learning-empowered wavefront shaping in nonstationary scattering media," *arXiv e-prints,* Accessed on: August 01, 2019 Available: https://ui.adsabs.harvard.edu/abs/2019arXiv190900210L

[4] I. M. Vellekoop and A. P. Mosk, "Focusing coherent light through opaque strongly scattering media," *Optics Letters,* vol. 32, no. 16, pp. 2309-2311, 2007/08/15 2007.

[5] L. V. Wang and S. Hu, "Photoacoustic Tomography: In Vivo Imaging from Organelles to Organs," *Science,* vol. 335, no. 6075, p. 1458, 2012.

[6] D. A. Boas *et al.*, "Imaging the body with diffuse optical tomography," *IEEE Signal Processing Magazine,* vol. 18, no. 6, pp. 57-75, 2001.

[7] T. Ni, L. Xie, H. Zheng, E. K. Fishman, and A. L. Yuille, "Elastic Boundary Projection for 3D Medical Imaging Segmentation," *arXiv e-prints,* Accessed on: December 01, 2018Available: https://ui.adsabs.harvard.edu/abs/2018arXiv181200518N

[8] R. Hu, P. Dollár, K. He, T. Darrell, and R. Girshick, "Learning to Segment Every Thing," *arXiv e-prints,* Accessed on: November 01, 2017Available: https://ui.adsabs.harvard.edu/abs/2017arXiv171110370H

[9] M. Berman, A. Rannen Triki, and M. B. Blaschko, "The Lovász-Softmax loss: A tractable surrogate for the optimization of the intersection-over-union measure in neural networks," *arXiv e-prints,* Accessed on: May 01, 2017Available: https://ui.adsabs.harvard.edu/abs/2017arXiv170508790B

[10] J. Chen, J. Chen, H. Chao, and M. Yang, "Image Blind Denoising with Generative Adversarial Network Based Noise Modeling," in *2018 IEEE/CVF Conference on Computer Vision and Pattern Recognition*, 2018, pp. 3155-3164.

[11] S. Lefkimmiatis, "Universal Denoising Networks : A Novel CNN Architecture for Image Denoising," *arXiv e-prints,* Accessed on: November 01, 2017Available: https://ui.adsabs.harvard.edu/abs/2017arXiv171107807L

[12] Y. Zhang, Y. Tian, Y. Kong, B. Zhong, and Y. Fu, "Residual Dense Network for Image Super-Resolution," *arXiv e-prints,* Accessed on: February 01, 2018Available: https://ui.adsabs.harvard.edu/abs/2018arXiv180208797Z

[13] C. Yang and F. Gao, "EDA-Net: Dense Aggregation of Deep and Shallow Information Achieves Quantitative Photoacoustic Blood Oxygenation Imaging Deep in Human Breast," in *Medical Image Computing and Computer Assisted Intervention – MICCAI 2019*, Cham, 2019, pp. 246-254: Springer International Publishing.

[14] L. Wang, C. Sun, Y. Fu, M. H. Kim, and H. Huang, "Hyperspectral Image Reconstruction Using a Deep Spatial-Spectral Prior," in *Proceedings of the IEEE Conference on Computer Vision and Pattern Recognition*, 2019, pp. 8032-8041.

[15] Y. Luo, S. Yan, H. Li, P. Lai, and Y. Zheng, "Deep learning assisted optical wavefront shaping in disordered medium," *International Society for Optics and Photonics*. SPIE, 2019.

[16] Y. Li, Y. Xue, and L. Tian, "Deep speckle correlation: a deep learning approach toward scalable imaging through scattering media," *Optica,* vol. 5, no. 10, pp. 1181-1190, 2018/10/20 2018.

[17] A. Turpin, I. Vishniakou, and J. d. Seelig, "Light scattering control in transmission and reflection with neural networks," *Optics Express,* vol. 26, no. 23, pp. 30911-30929, 2018/11/12 2018.

[18] T. Chaigne, O. Katz, A. C. Boccara, M. Fink, E. Bossy, and S. Gigan, "Controlling light in scattering media non-invasively using the photoacoustic transmission matrix," *Nature Photonics,* vol. 8, pp. 58-64, January 01, 2014 2014.

[19] S. M. Popoff, G. Lerosey, R. Carminati, M. Fink, A. C. Boccara, and S. Gigan, "Measuring the Transmission Matrix in Optics: An Approach to the Study and Control of Light Propagation in Disordered Media," *Physical Review Letters,* vol. 104, no. 10, p. 100601, 03/08/ 2010.

[20] J. V. Thompson, B. H. Hokr, and V. V. Yakovlev, "Optimization of focusing through scattering media using the continuous sequential algorithm," (in eng), *Journal of modern optics,* vol. 63, no. 1, pp. 80-84, 2016.

[21] D. B. Conkey, A. N. Brown, A. M. Caravaca-Aguirre, and R. Piestun, "Genetic algorithm optimization for focusing through turbid media in noisy environments," *Optics Express,* vol. 20, no. 5, pp. 4840-4849, 2012/02/27 2012.

[22] P. Vincent, H. Larochelle, I. Lajoie, Y. Bengio, and P.-A. J. J. o. m. l. r. Manzagol, "Stacked denoising autoencoders: Learning useful representations in a deep network with a local denoising criterion," *Journal of machine learning research*, vol. 11, no. Dec, pp. 3371-3408, 2010.

[23] D. P. Kingma and M. Welling, "Auto-Encoding Variational Bayes," *arXiv e-prints,* Accessed on: December 01, 2013Available: https://ui.adsabs.harvard.edu/abs/2013arXiv1312.6114K

[24] I. J. Goodfellow *et al.*, "Generative Adversarial Networks," *arXiv e-prints,* Accessed on: June 01, 2014Available: https://ui.adsabs.harvard.edu/abs/2014arXiv1406.2661G

[25] L. A. Gatys, A. S. Ecker, and M. Bethge, "A Neural Algorithm of Artistic Style," *arXiv e-prints,* Accessed on: August 01, 2015Available: https://ui.adsabs.harvard.edu/abs/2015arXiv150806576G

[26] P. Isola, J.-Y. Zhu, T. Zhou, and A. A. Efros, "Image-to-Image Translation with Conditional Adversarial Networks," *arXiv e-prints,* Accessed on: November 01, 2016Available: https://ui.adsabs.harvard.edu/abs/2016arXiv161107004I

[27] A. Boesen Lindbo Larsen, S. Kaae Sønderby, H. Larochelle, and O. Winther, "Autoencoding beyond pixels using a learned similarity metric," *arXiv e-prints,* Accessed on: December 01, 2015Available: https://ui.adsabs.harvard.edu/abs/2015arXiv151209300B

[28] O. Ronneberger, P. Fischer, and T. Brox, "U-Net: Convolutional Networks for Biomedical Image Segmentation," *arXiv e-prints,* Accessed on: May 01, 2015Available: https://ui.adsabs.harvard.edu/abs/2015arXiv150504597R

[29] Y. Luo, S. Yan, H. Li, P. Lai, and Y. Zheng, "Focusing light through scattering media by reinforced hybrid algorithms," *APL photonics,* vol. 5, no. 1, 016109, 2020.

[30] I. M. Vellekoop and A. P. Mosk, "Phase control algorithms for focusing light through turbid media," *Optics Communications,* vol. 281, no. 11, pp. 3071-3080, 2008/06/01/ 2008.